\documentclass[aps,prl,twocolumn,groupedaddress]{revtex4}
\usepackage{color}
\usepackage{graphics}
\usepackage{epsfig}
\usepackage[outercaption]{sidecap}

\newcommand{\bee}{\begin{equation}}
\newcommand{\ene}{\end{equation}}
\newcommand{\beea}{\begin{eqnarray}}
\newcommand{\enea}{\end{eqnarray}}

\begin{document}
\title{Evidence of new finite beam plasma instability for magnetic field generation}
\author{Amita Das$^1$}
\email{amita@ipr.res.in}
\author{Atul Kumar$^1$}
\author{Chandrasekhar Shukla$^1$}
\author{Ratan Kumar Bera$^1$}
\author{Deepa Verma$^1$} 
\author{Bhavesh Patel$^2$}
\author{Y. Hayashi$^3$}
\author{K. A. Tanaka$^4$}
\author{Amit D. Lad,$^{5}$}
\author{G. R. Kumar$^5$}
\author{Predhiman Kaw$^1$} 
\affiliation{$^1$Institute for Plasma Research, HBNI,  Bhat, Gandhinagar - 382428, India }
\affiliation{$^2$Institute for Plasma Research, Bhat, Gandhinagar - 382428, India }
\affiliation{$^{3}$ Graduate School of Engineering, Osaka University, Suita, Osaka 565-0871 Japan }
\affiliation{$^{4}$  Extreme Laser Infrastructure-Nuclear Physics 30 Reactorului, PO Box MG-6, Bucharest Magurele 077125 Romania} 
\affiliation{$^{5}$ Tata Institute of Fundamental Research, 1 Homi Bhabha Road, Mumbai 400005, India} 

\begin{abstract} 
{\bf We demonstrate by computer simulations, laser plasma experiments and analytic theory that a hitherto unknown 
 instability  is excited in  the beam plasma system with finite transverse size. This instability is responsible for the generation of 
  magnetic fields at scales comparable to the transverse beam dimension which  can be much longer than the electron skin depth 
scale.    This counterintuitive result 
 arises due to  radiative leakage associated with finite beam boundaries which are  absent in conventional  infinite periodic systems 
 considered in earlier simulations as well as theoretical analyses and may trigger a reexamination of hitherto prevalent  idea.}

\end{abstract}
\maketitle 

The dynamical evolution of intense magnetic fields and associated current pulses  plays an important role in a variety of plasma physics problems. 
These include   plasma switches,  designing of novel 
radiation and charged particle sources, laser driven fusion and also laboratory simulation of astrophysical phenomena etc \cite{PhysRevLett.81.822, RevModPhys.46.325, Betti2016, Mondal, Remington,Gaurab_2017}. The current pulses may be 
 generated and driven into a solid target by an intense,
femtosecond laser that generates a hot, dense plasma at the surface.  It is now well known that  giant current pulses 
produced in such an interaction induce return currents from the  thermal 
plasma electrons and these two types of currents are subject to electromagnetic instabilities. It is widely believed that the well known Weibel instability \cite{Weibel, Pegoraro1996}  separates these  initially superimposed  
counter currents  and leads to giant magnetic field generation in intense laser driven plasmas. Countless analytical and simulation studies have explored this instability 
and predicted  that  the scale at  which the magnetic field gets generated characterizes this instability and is essentially the  
(local) skin depth of the plasma. The nonlinear inverse cascade mechanism is  invoked for subsequent long scale generation of magnetic fields. 
 This belief had strong support from   
 all plasma simulations which demonstrated unambiguously  that  the scale  at which the energy is driven into the magnetic field  is  at the skin depth scale. 
 In this manuscript we provide convincing  evidence that this is  not the 
correct physical situation but  is merely an artifact of boundless beam plasma interaction in simulations as well as theoretical analyses. 
We demonstrate through laser plasma experiments and  finite 
beam size   simulations 
using different variety of codes, that the energy is input preferentially into the plasma at the  scale size of the beam 
(in the case of laser experiments this is at  the spot size of the irradiating 
laser).  We show that magnetic field 
generation caused by the electron currents first occurs at the scale size of the boundary and only much later does the Weibel instability kick in at the skin depth scale. 
This is a new instability mechanism 
of   magnetic field generation   that may  tentatively be called   'finite beam  instability' (FBI).  It has been demonstrated with the help of a 
detailed linearized perturbation analysis that the 
 characteristic features of FBI  are 
in marked contrast with Weibel instability. Such a behaviour finds 
support in   our finite beam size simulations. 
We  show that the FBI dominates the Weibel instability at the initial stage and  
has its origin in the effect of radiative loss from   negative energy waves  in 
 a manner analogous to the radiative instability of  a leaky  waveguide \cite{Miyagi, Botez}.  We believe our study is an important step in establishing  the crucial 
role that  finite size effects can play in qualitatively changing the physical nature of eigen modes in beam plasma system.

The choice of infinite beam - plasma periodic system considered in earlier studies \cite{PhysRevE.65.046408} 
is based on  an inherent assumption  that the boundary effects due to the
finite system would merely have a small incremental impact. This assumption, however, turns out to be incorrect.  
 We show with   PIC as well as two-fluid simulations that when a beam with finite 
 transverse extent is considered,  an  entirely new instability associated with the finite size of the beam  
appears which generates  magnetic fields at the scale of the beam size right from the very beginning. 
Secondly,  the  sheared electron flow configuration at the two edges of the finite beam 
is seen to be susceptible to Kelvin Helmholtz instability \cite{EMHD, cshukla}. Weibel appears only in the bulk 
region of the beam at a later stage. 
Thus, there are three  sources of magnetic fluctuations in a finite beam system - the new Finite 
Beam Instability (FBI) and KH instability
operating at the edge and the usual  Weibel destabilization process occurring in the  bulk 
region.  The theoretical description of the FBI  has been provided here which is followed 
up by evidences from simulations (both PIC and fluid [LCPFCT\cite{LCPFCT}]) and experimental data\cite{Mondal} where the 
appearance of magnetic field at the laser spot size (much larger than the skin depth) at very early stage can only be accounted via this instability. 
The  conventional Weibel destabilization route of the beam plasma system can never 
account for such an effect.

An  equilibrium configuration  of the beam plasma system in 2-D $x-y$ plane is 
considered as shown in Fig.1.
The central region II from $-a \le y \le a$ carries the beam 
current and an oppositely flowing background plasma current,  which balances each other. In region 
I and region III the plasma is static 
and at rest. 
The  charge  neutralization in equilibrium is achieved  
 by  balancing the total electron density by the 
 background ion density, viz., $\sum_{\alpha} n_{0\alpha} = n_{0i}$ in all the three regions.  
 The  electron flow velocity in region I and III is zero, whereas in region II it 
satisfies the condition of zero current, i.e. 
$ \sum_{\alpha} n_{0\alpha} v_{0\alpha x} = 0$. Here the suffix $\alpha $ stands for  
 $b$ and $p$ representing the     beam and plasma 
 electrons. 
The linearized perturbation of this equilibrium is considered with  variations in 
 $y$ and $t$ only. The flow is confined in $x- y$ plane, so  we have
 $B_{1z}$, $E_{1x}$ and $E_{1y}$ (where the suffix $x,y,$ and $z$ denotes 
 the components and B and E represents the magnetic and electic fields) only as  the perturbed dominant fields.  
Eliminating all the perturbed fields in terms of $E_{1x}$ leads to the  following differential 
equation:  
\begin{equation}
\left[f_2 E_{1x}^{\prime} \right]^{\prime} - g_2 E_{1x} = 0
\label{finaleq}
\end{equation}
Here 
\begin{eqnarray}
f_2 &=& 1 + \frac{S_4}{\omega^2} - \frac{S_3^2}{\omega^2(S_1-\omega^2)} \\
g_2 &=& S_2 - \omega^2
\end{eqnarray}
where 
\begin{eqnarray}
S_1 &=& \sum_{\alpha} \frac{n_{0 \alpha}}{n_0 \gamma_{0 \alpha}}; \\
S_2 &=& \sum_{\alpha} \frac{n_{0 \alpha}}{n_0 \gamma_{0 \alpha}^3};\\
S_3 &=& \sum_{\alpha} \frac{n_{0 \alpha} v_{0 x \alpha}}{n_0 \gamma_{0 \alpha}}; \\
S_4 &=& \sum_{\alpha} \frac{n_{0 \alpha} v_{0 x \alpha}^2}{n_0 \gamma_{0 \alpha}};
\label{sdefs}
\end{eqnarray}
It should be noted that $S_3$ and $S_4$ are finite only when there is an equilibrium flow in the 
two fluid electron depiction. 
Furthermore, if the flow velocities of the two electron species are equal and opposite  then $S_3=0$.

The homogeneous limit of Califano $ \emph{et. al.} $\cite{PhysRevE.58.7837} can be easily  recovered if we 
take Fourier transform of Eq.(\ref{finaleq}). The homogenous equation yields the dispersion relation 
for the  Weibel growth rate. 
We now seek the possibilities for obtaining purely growing modes in a finite system. 
For this purpose we multiply  Eq.(\ref{finaleq}) by $E_{1x}$, replace $\omega^2 = -\gamma^2$ (for 
purely growing modes) and integrate over $y$ over region II, i.e. from $-a$ to $a$. This yields: 
 \begin{equation}
\int_{-a}^a \left[ E_{1x} (f_2 E_{1x}^{\prime})^{\prime} - g_2 E_{1x}^2\right]dy = 0  \\
\label{inteq}
 \end{equation}
 Upon integrating by parts we obtain 
 \begin{equation}
f_2 \left[ E_{1x} E_{1x}^\prime \right]\vert_{-a}^{a} - \int_{-a}^{a} \left\lbrace f_2 \left[ E_{1x}^{\prime}\right]^2 +g_2 
E_{1x}^2 \right\rbrace dy = 0 \\
\label{inteq1}
 \end{equation}
In region II, $f_2$ and $g_2$ being constant, we can take them outside the integral. Thus Eq.(\ref{inteq1}) 
can be written as 
\begin{equation} 
f_2 \left[ E_{1x} E_{1x}^\prime \right]\vert_{-a}^{a} - f_2 \int_{-a}^{a}  \left[ 
E_{1x}^{\prime}\right]^2 dy  - g_2 
\int_{-a}^{a}E_{1x}^2  dy = 0 \\
\label{inteq2}
\end{equation}
If the boundary term is  absent, as in the case of infinite homogeneous system, then Eq.(\ref{inteq2}) 
can be 
satisfied for a finite value of $E_{1x}$, provided second and third terms have opposite signs. 
The integrand being positive definite this is possible provided $g_2$ and $f_2$ have opposite signs. 
The definitions of $g_2$ and $f_2 $ in terms of $\gamma^2$ are 
$$ f_2 = 1 + \frac{S_3^2}{\gamma^2(S_1+\gamma^2)} - \frac{S_4}{\gamma^2} $$
$$ g_2 = S_2 + \gamma^2$$
Since $g_2$ is  positive,  the only possible way for $f_2$ to be negative is to have 
$S_4/\gamma^2$ dominate over the first two terms of $f_2$. Thus, the conventional Weibel 
gets driven by $S_4$. It is also obvious that $S_3$ provides a stabilizing contribution making it 
more difficult for $f_2$ to become negative. A finite value of $S_3$ implies a non-symmetric flow 
configuration,  i.e. one for which the two  electron species have different flow speeds.  

For finite system, something interesting happens when  boundary contribution are retained.  
The value of $\left[ E_{1x} E_{1x}^\prime \right]\vert_{-a}^{a}$ should be positive as 
 $E_{1x}^2$ should increase as one enters region II from region I (at $y = -a$) and it should decrease 
 at $y = a$. Thus the sign of first term will be  determined by the sign of $f_2$. 
 Another way to understand the positivity of  the sign of 
 $\left[ E_{1x} E_{1x}^\prime \right]\vert_{-a}^{a}$ 
 is by realizing that this term is essentially the radiative flux moving outside region II
 which can only be positive. This is seen  by casting it  in the form  of the Poynting flux by expressing the 
 derivative of $E_{1x}$ in terms of 
 $B_{1z}$.

There exists the possibility then  that the first term 
of Eq.(\ref{inteq2}) has a finite contribution to balance the second and the third terms. 
Thus even if $f_2$ and $g_2$ have same signs (positive) Eq.(\ref{inteq2}) 
can be satisfied for a finite $E_{1x}$ if the boundary is finite and boundary terms contribute through the Poynting flux. It should be noted that the  
instability driven in this case is 
different from the Weibel mode as the boundary terms are responsible for it and $S_3$ is playing an altogether different role of 
destabilization. 
Furthermore, it is interesting to observe  that Equation(\ref{inteq2}) can be 
satisfied most easily by the boundary contribution provided the variations in $E_{1x}$ in the bulk is 
minimal 
so as to have minimal (close to zero) contribution from the second term. Thus 
the instability driven by the boundary term would have a preference 
for long scale excitation. 

\begin{figure*}[h!]
\center
                \includegraphics[width=\textwidth]{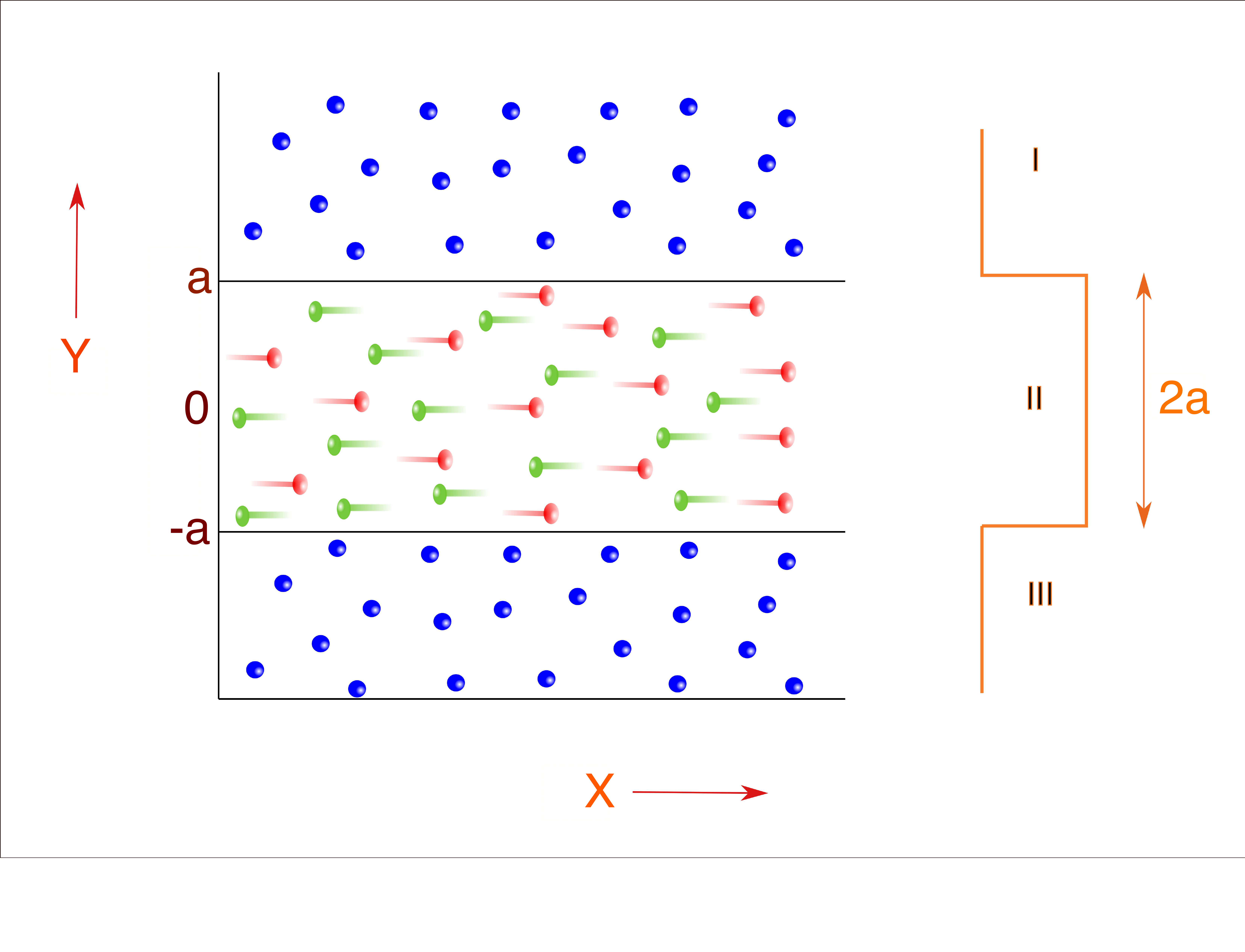} 
             \caption{ Schematics of 2D- equilibrium geometry of the beam plasma system where beam has finite width '$ 2a $' in the transverse direction   }  
                 \label{fig1}
         \end{figure*}
\begin{figure*}[h!]
  \center
                \includegraphics[width=\textwidth]{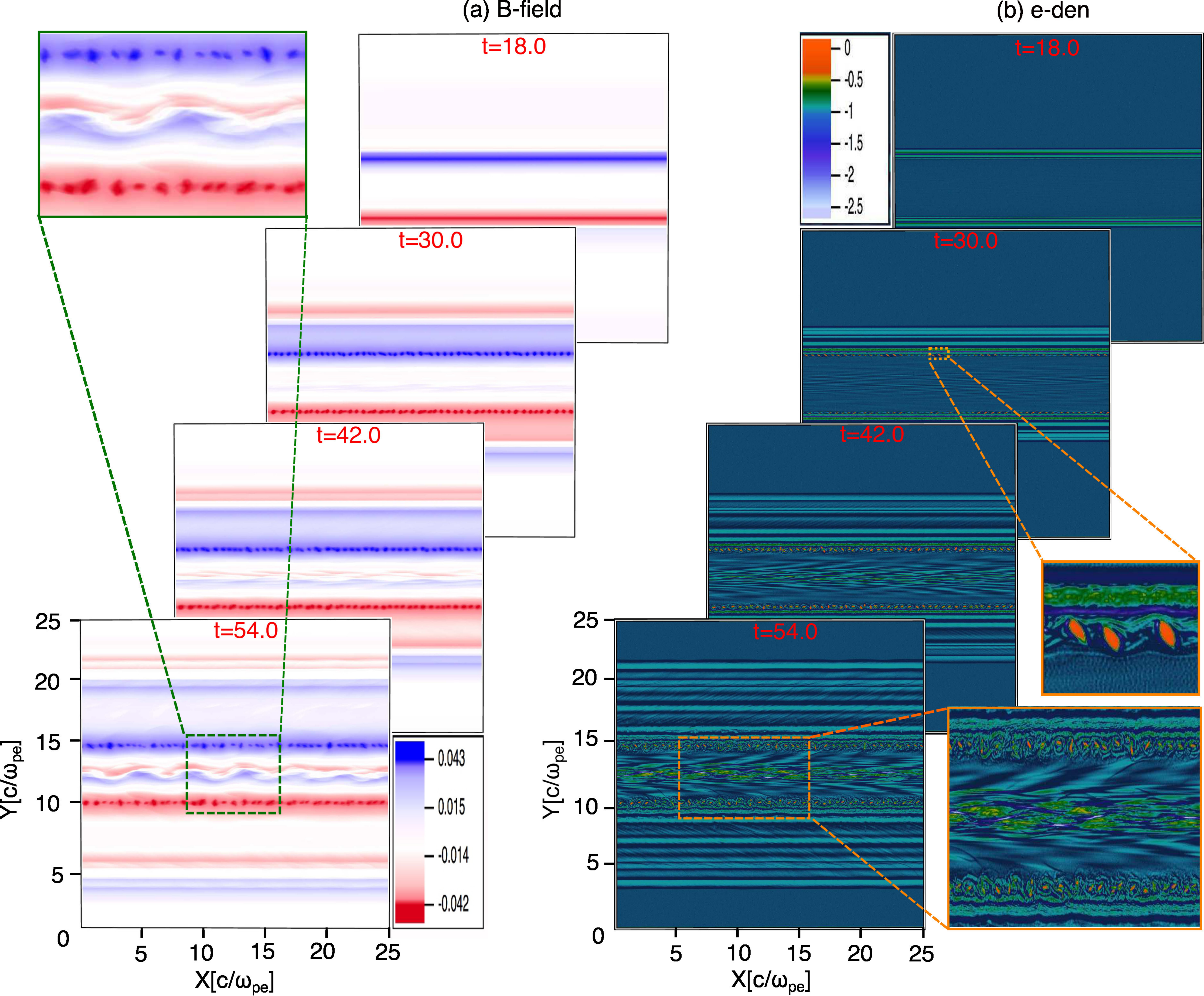} 
               \caption{ Snapshots of  magnetic field B [in the units of $(m c\omega_{pe}/e)$] and electron density evolution in  time t [in the unit of $\omega_{pe}^{-1}$] 
               has been depicted [prepared by Vapor \cite{clyne2007interactive, clyne2005prototype}]. Case(a) and Case(b) depicts the spatio temporal evolution of  magnetic field $ B_z $ and the electron density respectively. At time $t=18.0$ only FBI development ($B_z$ field with the opposite polarity) can be seen, at $t=30$ KH mode at the beam edge  and a faint Weibel 
               instability in the bulk  appears. At $t = 42.0$ and $t=54.0$ all the three instabilities can be observed clearly. }  
                 \label{fig2}
         \end{figure*}

%
PIC simulations using OSIRIS \cite{Fonseca2002, osiris}, PICPSI \cite{Shukla2015} and EPOCH \cite{epoch} were carried out for the case of a 
forward beam current and a compensating return plasma current 
of  a finite  transverse extent  at $t = 0$ shown as the  equilibrium configuration in Fig.(\ref{fig1}). 
The simulations were carried out in both 2-D and 3-D. In 3-D compensating cylindrical beam currents
 of diameter $2a$ was chosen. 
The   2-D simulations were carried for a   box size of 
$25 c/\omega_{pe} \times 25 c/\omega_{pe}$.  The beam was chosen to be  confined within 
an extent of $ 2a = 5 c/ \omega_{pe}$. Various choices of beam and background electron 
density have been chosen. In Fig.2  we have chosen to show the results (evolution of magnetic field as well as density)  for 
   beam electron density of $0.1 n_0$ moving  along $\hat{x}$ with a velocity of $0.9c$ 
   in a central region( from  $y = 10c/\omega_{pe}$ to $y = 15 c/\omega_{pe}$ ) 
i.e., with   a transverse extent of 
 $5 c/\omega_{pe}$. In the same region, a  shielding return current along $-\hat{x}$ of background electrons with density $0.9n_0$ has been taken to flow 
 with a velocity of $0.1c$.   
  Here,  $n_0$ is the density of background ions which are at rest everywhere.  In the remaining region  from $y = 0$ to $10c/\omega_{pe}$ and $15$ to $20$ 
  $c/\omega_{pe}$
   electrons and ions both with density $n_0$ are at rest. Thus, the plasma everywhere is neutral with electron density balancing the density of 
  background plasma ions.  In the central beam region, the beam current is exactly compensated by the return shielding current. 
  The movie uploaded as supplementary material shows the generation and evolution magnetic field for  this particular configuration.  
   Snapshots of the evolution at various 
  times have been depicted  in Fig.~\ref{fig2} in the form of 2-D color plots  for the $z$ component of the magnetic field and the perturbed charge density respectively.

\begin{figure*}[h!]
  \center
                \includegraphics[width=\textwidth]{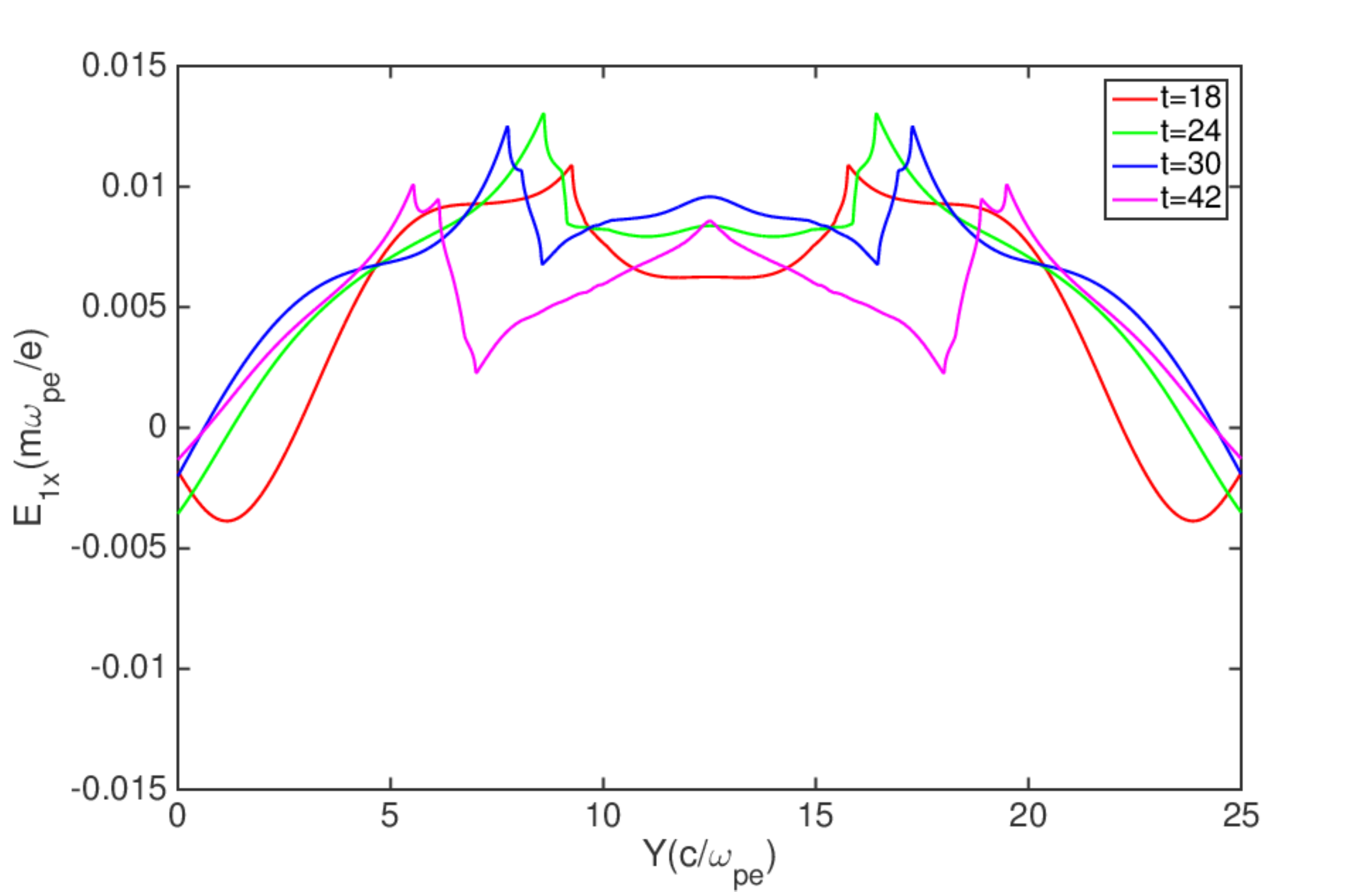} 
              \caption{Evolution of $E_{1x}$  profile as a function of $y$ showing the minimal variation inside the beam region in the beginning in confirmity with 
              FBI mode. }  
                 \label{fig4}
         \end{figure*}

  From these plots, it is clearly evident that there are  three distinct phases of evolution. 
  During the first phase 
 from $t = 0.12 {\omega_{pe}}^{-1}$ to $t = 30 {\omega_{pe}}^{-1} $ perturbed $z$ component of 
 magnetic field along $\hat{z}$ (transverse to 
 both flow and inhomogeneity)   appear  at the edges with 
  opposite polarity. This magnetic field has no $x$ dependence and is a function of $y$ alone. 
  The magnetic field perturbations are seen to grow with time and also 
  expand in $y$ from the edges in both the directions  at the speed of light. 
  The electron density perturbations, which also appear 
  at the edge,  on the other hand,  remain confined at the edge  during this phase. 
    This  first phase of the evolution thus can   be characterized by the appearance 
     of  magnetic field perturbations with variations only along  $\hat{y}$, the transverse 
     direction. This fits the analytical description very well. Keeping in view that  
     the structures do not seem to vary with respect to the $x$,  the 1-D profiles along 
     $y$ have been shown in Fig.~\ref{fig4} for $E_{1x}$. 
     As predicted for the FB mode theoretically,  $E_{1x}$ shows minimal variation inside the beam 
     region. The PIC simulations were also repeated for the case where $S_3 =0$ was taken 
     by a choice of symmetric flow. In this case we observed that the FB mode did not appear. 
     This shows that $S_3$ plays a destabilizing role for this instability, just the opposite role it has for the  
      Weibel mode. During the second phase from $t = 30 {\omega_{pe}}^{-1} $  the Kelvin Helmholtz (KH) like 
perturbations 
appear at the edge of the current. Around this time 
one can also observe appearance of faint Weibel perturbations in the bulk central region.  Both the KH and the Weibel mode 
have variations   along both $\hat{y}$ and $\hat{x}$ directions.  

\begin{figure*}[h!]
  \center

                \includegraphics[width=\textwidth]{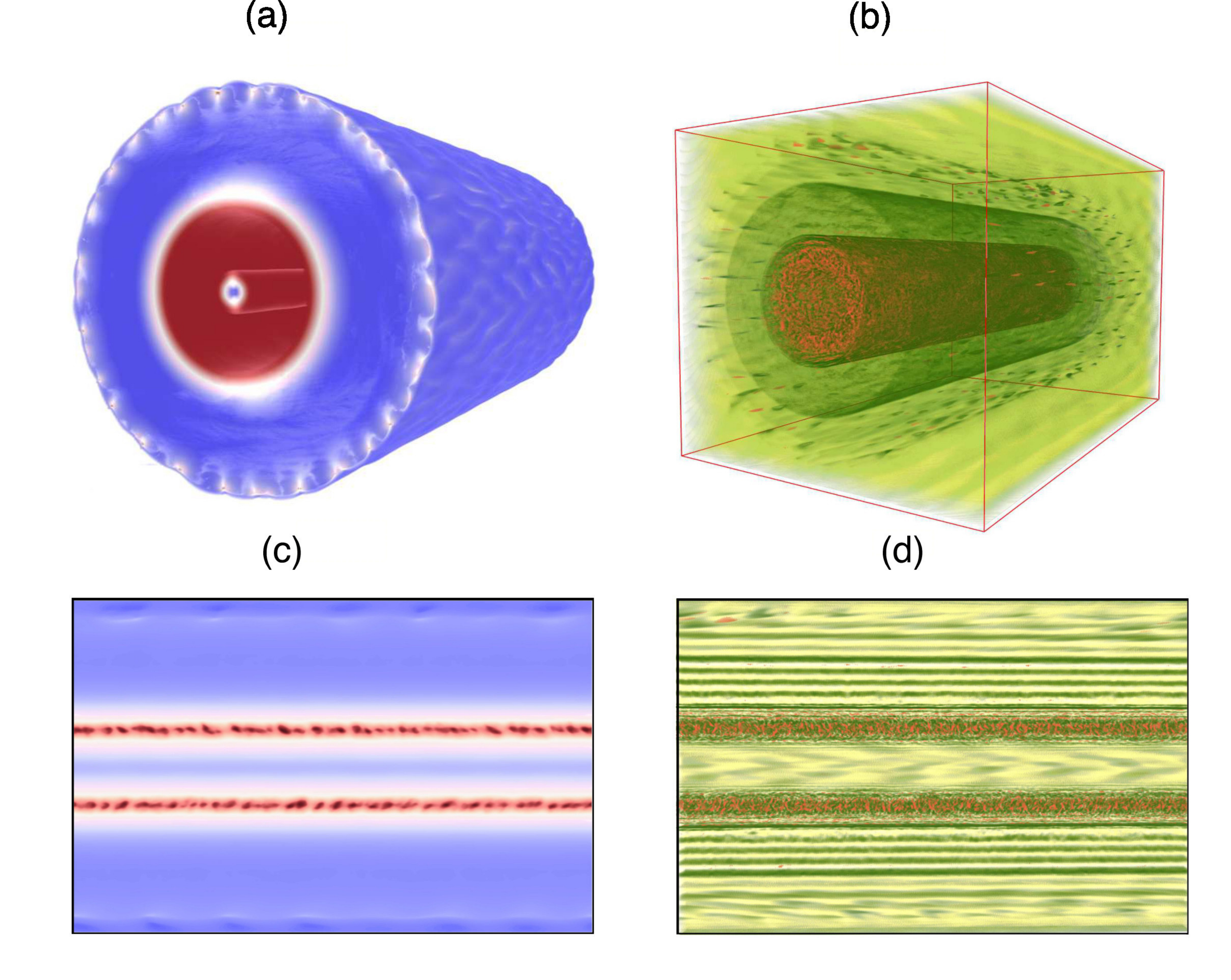} 
              \caption{ Snapshots of 3D PIC simulation with OSIRIS  [prepared by Vapor \cite{clyne2007interactive, clyne2005prototype}] at  time $ t=78.6 $ 
               in the unit of $\omega_{pe}^{-1}$. Subplots (a) and (b) depict  3D volume rendering of poloidal magnetic field and electron density respectively. 
               In subplot (c) and (d) the cross sectional view in the $r - z$ plane have been shown for the magnetic field and the electron density respectively. 
               In  these 3-D snapshots also all the three instabilities viz., FBI and KH at the edge of the beam and Weibel in the bulk region of the beam  are clearly evident. }
                 \label{fig6}
         \end{figure*}

We also provide the snap shot  from 3-D simulations in Fig. (\ref{fig6}) 
for a finite beam propagating in the 
plasma at a very late time $t = 78.6 {\omega_{pe}}^{-1}$ 
in which both FBI and the Weibel mode can be observed. 
These observations with characteristics three  phase 
developments have been  repeatedly observed in both 2-D and 3-D  from a variety 
of  simulations carried out with different PIC  as well as fluid codes.  

\begin{figure*}[h!]
  \center
                \includegraphics[width=\textwidth]{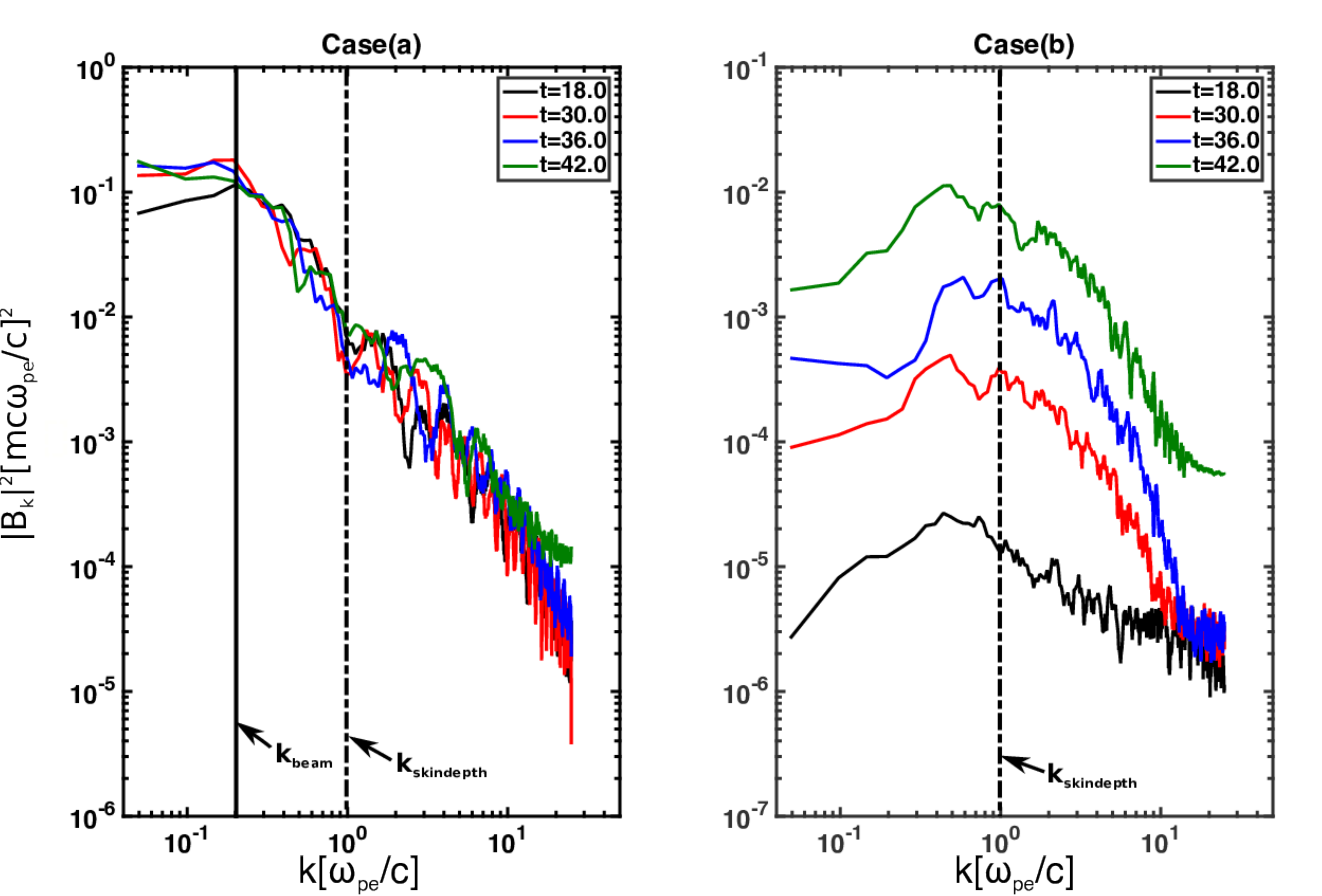} 
              \caption{Evolution of Magnetic field spectra from PIC simulations with OSIRIS Case(a) Finite beam-plasma system 
              was considered in simulation. The spectral maximum is at  at the width of the beam right from the beginning; Case(b)-Periodic box simulations corresponding to 
              infinite beam-plasma system where peak of the field spectra appears at the  electron skin depth due to  Weibel mode and the nonlinear cascade towards long scale 
              can be seen to be very slow. }  
                 \label{fig5}
         \end{figure*}

A comparison of the  
 evolution of the magnetic field spectra for the periodic infinite system as well as 
 the finite beam case has been provided 
 at various times  in Fig.~\ref{fig5}. It is clear from the figure that for the periodic 
 infinite case the peak of  spectral power appears at the electron skin depth scale initially. 
 In this case,  the  spectral power only subsequently  cascades towards longer scales 
 via inverse  cascade mechanism which is considerably slow.  On the other hand,  for the finite case, it can be clearly observed 
 that the spectral peak appears at the beam size right from the very beginning. 

\begin{figure*}[h!]
  \center
                \includegraphics[width=\textwidth]{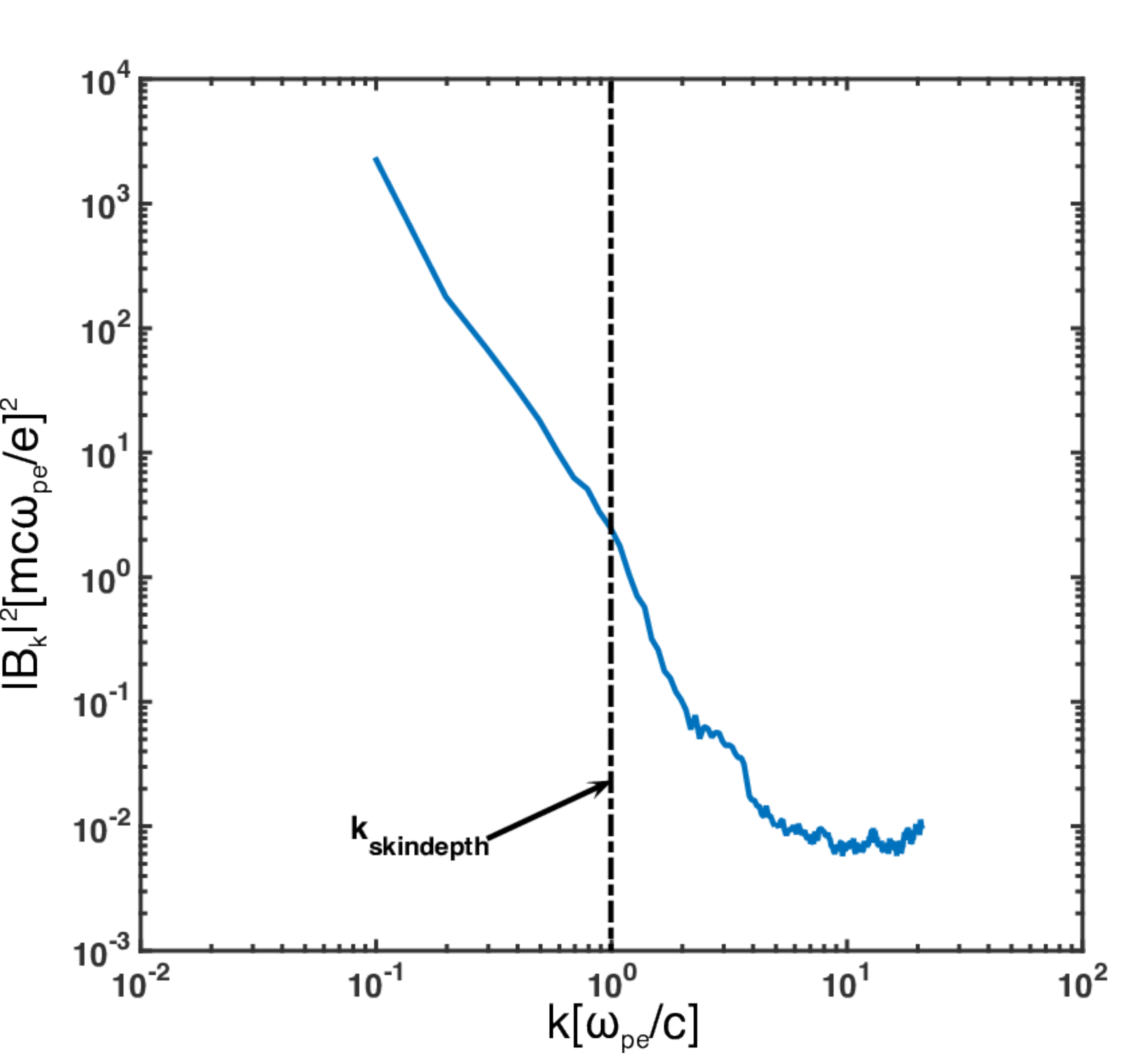} 
              \caption{Power spectrum of magnetic field spatial profile measured at pump-probe delay of $ 0.4  $ picoseconds}  
                 \label{fig7}
         \end{figure*}

 In laboratory laser plasma experiments  the electron beam width would be 
 finite, typically commensurate with the dimension of the laser focal spot.  It would, therefore, be interesting to look for evidences 
 of finite beam instability in the laser plasma experiments. 
 The experiments were carried out at the Tata Institute of Fundamental Research, Mumbai using a 20 TW Ti: Sapphire laser. 
 The laser is capable of delivering 30 fs, 800 nm pulses at 10 Hz. A p-polarized pump laser with 120 mJ energy was focussed on an aluminium coated BK7 glass target 
 with an $f/3$ off-axis parabola. The resulting peak intensity was $3 \times  10^{18} W/cm^2$. A second harmonic ($400 nm$) probe pulse was time delayed with respect to the pump 
 pulse. The intensity of probe pulse was adjusted to be $~10^{11} W/cm^2$. A second harmonic probe can penetrate up to four times the critical density corresponding
 $800 nm$ pum radiation\cite{Gaurab}. A magneto-optic Cotton-Mouton polarimetry set-up \cite{Mondal,Gaurab} was utilized to spatially resolve ellipticity. 
 The induced magnetic field is inferred from the ellipticity data \cite{sandhu2002}. 
 The power spectra have been calculated as per Ref. \cite{Mondal} and shown to peak at the focal spot of the laser pulse 
 right from the very begining. In  Fig.~(\ref{fig7})  observations at a  time delay of $0.4$ pico-seconds have been shown. 
It is clear from the figure  that  the longest observed scale of focal spot has the maximum power at such an early time. 
 Thus,  this  experiment provides another evidence for the existence of  the FBI instability.

While the KH and the Weibel modes are well known and have been discussed extensively in 
the literature, the FBI  has neither been observed nor been described anywhere. This qualitatively new result arises because the infinite periodic approximation is unable to consider 
effects due to radiative leakage at the boundaries. The implications of this particular instability on magnetic field generation 
need to be evaluated in different contexts. For instance, it is likely that the 
finite size  jets emanating from astrophysical objects are susceptible to  
this particular instability.  This work also suggests that the finite size considerations 
in many other systems need to be looked afresh to unravel new effects which might 
have been overlooked so far. 

 
Finally, some general remarks. Homogeneity and infinite extent are   idealizations that permeate all physical models as they simplify descriptions because of  the resulting uniformity and reduced 
dimensionality of the problem.  Real systems, however,  are finite and  deviations from idealization can lead to novel physical effects. This lesson has been learned from time to time. For instance, the existence of Casimir effect \cite{PhysRev.73.360,PKNAW} leading to attractive force between two plates due to vacuum fluctuations, 
 einzel lense \cite{Liebl_2008, Feynmann} - which enables focussing of charged particles using fringing fields at capacitor edges, existence of surface plasmon modes \cite{PhysRev.120.130} distinct from bulk modes, 
discrete eigen modes in waveguides etc., are some examples of finite size boundaries leading to new effects. 
  The present manuscript demonstrates another outstanding example in this context. 

%
%
%
%
\section*{Acknowledgments}
AD acknowledges the DAE-SRC-ORI grant Government of India. GRK acknowledges J C Bose Fellowship grant (JCB-037/2010) from the Department of Science and Technology, Government of India. A part of the research has also been supported by the S-level funding 15H05751, JSPS Japan.


\bibliographystyle{ieeetr}  

\bibliography{finite_arxiv}

\end{document}